\documentclass[fleqn,usenatbib]{mnras}

\usepackage{xcolor}

\usepackage{newtxtext,newtxmath}

\usepackage[T1]{fontenc}

\DeclareRobustCommand{\VAN}[3]{#2}
\let\VANthebibliography\thebibliography
\def\thebibliography{\DeclareRobustCommand{\VAN}[3]{##3}\VANthebibliography}

\usepackage{graphicx}	
\usepackage{amsmath}	
\usepackage{commath}
\usepackage{subcaption}
\usepackage{orcidlink}

\newcommand{\der}{\mathrm{d}}

\title[Darkness Visible]{Darkness Visible:\\N-Body Simulations of Dark Matter Spikes in Hernquist Haloes}

\author[Kamermans \& Wierda]{
Jasper Leonora P. D. Kamermans \orcidlink{0000-0001-7003-7237}$^{1}$
A. Renske A. C. Wierda \orcidlink{0000-0002-9569-2745}$^{2}$\thanks{E-mail: wierda@kth.se}
\\
$^{1}$Department of History and Art History, Universiteit Utrecht, Heidelberglaan 8, 3584 CS Utrecht, The Netherlands\\
$^{2}$Department of Physics, KTH Royal Institute of Technology, The Oskar Klein Centre, AlbaNova, SE-106 91 Stockholm, Sweden}

\date{Accepted XXX. Received YYY; in original form ZZZ}

\pubyear{\the\year{}}

\begin{document}
\label{firstpage}
\pagerange{\pageref{firstpage}--\pageref{lastpage}}
\maketitle

\begin{abstract}
Dark matter is theorised to form massive haloes, which could be further condensed into so-called spikes when a black hole grows at the centre of such a halo. The existence of these spikes is instrumental for several dark matter detection schemes such as indirect detection and imprints on gravitational wave inspirals, but all previous work on their formation has been (semi-)analytical. We present fully numerically simulated cold dark matter spikes using the SWIFT code. Based on these results, we propose a simple empirical density profile - dependent on only a single mass-ratio parameter between the black hole and total mass - for dark matter spikes grown in Hernquist profiles. We find that the radius of the spike scales differently compared to theoretical predictions, and show a depletion of the outer halo that is significant for high mass-ratio systems. We critically assess approximations of the spike as used in the field, show that our profile significantly deviates, and contextualise the potential influence for future DM detections by simulating binary black hole inspirals embedded in our profile.
\end{abstract}




\section{Introduction}
One of the biggest mysteries in modern cosmology is the origin of Dark Matter (DM). While never directly detected, it has been indirectly observed in systems ranging from small galaxies to clusters of galaxies, with more work underway focused on direct detection or solving theoretical challenges (for an overview, see \citealt{PDG}). The most recent \textit{Planck} satellite survey found, assuming standard $\Lambda$CDM cosmology, that $26.1\%$ of the total mass-energy of the universe consists of this DM, which is $83.9\%$ of the total matter density \citep{Planck:2018vyg}. DM is theorised to form small density perturbations in the early universe, which stop expanding after sufficient growth and collapse into a spherical halo \citep*{NFW,Bullock:2017xww}.\footnote{If this collapse is too fast, a BH forms instead \citep{Green:2020jor}.} While this work focuses on collisionless Cold DM, as it is part of $\Lambda$CDM cosmology, many other forms of DM have been proposed in order to solve known conflicts with this cosmology (see e.g. \citealt{Bullock:2017xww}). 

Several methods of detecting DM have been proposed and tried out: from direct detection in a laboratory to indirect detection using signals from astrophysical systems \citep{PDG}. One of these systems is a DM halo with an embedded BH at the centre. For potential detections, the density distributions of these systems must be well understood. This article will numerically simulate DM haloes with a central BH, and contextualise the results using the possible detection of DM through gravitational waves (GWs). There are more detection methods where the halo profile is of great importance, such as the detection of gamma-rays from DM self-annihilation (e.g. \citealt{Gondolo,BertoneGammaRaysIMBH, gammaray}), but this work will only consider the impact on the GW interpretation. 

\citet{Eda:2013gg} suggested that DM haloes slow down the inspiral of a compact object into an embedded BH. This leads to a dephasing of the generated GWs compared to the vacuum solution, which could be measurable with the next generation of gravitational wave detectors (e.g. \citealt{Eda:2014kra,Yue:2018vtk, Kavanagh:2020cfn, Coogan:2021uqv}). Recent numerical studies found that this dephasing is both larger than previously predicted, and the DM halo is depleted at a slower rate than anticipated, reinforcing the potential of this method \citep{Mukherjee:2023lzn,Kavanagh:2024lgq}. Future GW detectors such as the LISA gravitational wave detector should not only be able to measure this dephasing, but also differentiate between DM haloes and other environments such as accretion discs \citep{LISAreport,Cole:2022yzw}.

LISA will be able to detect extreme and intermediate mass-ratio inspirals where the central BH has a mass between $10^3$ $\text{M}_\odot$ and $10^6$ $\text{M}_\odot$ \citep{LISAreport}. These intermediate mass BHs (IMBHs) have been detected in our universe (e.g. \citealt{Farrell:2009uxm,Pasham:2014ybe,2024Natur.631..285H}), and there exists a catalogue of candidates from low-luminosity active galactic nuclei \citep{Barrows_2019}. The formation of these IMBHs at galactic centres likely mirrors that of supermassive BHs (SMBHs), though having experienced very few major galactic mergers, leaving the surrounding DM halo relatively unperturbed.

It has been known for some time that the growth of a BH influences the distribution of matter around it (e.g. \citealt{Quinlan:1994ed}). \citet{Gondolo} (From here on: G\&S) predicted that a growing BH in a DM halo creates a DM \textit{spike}, where the DM is redistributed into a steep cusp of uniform slope in the central regions. They analytically showed that if this growth is adiabatic, an initial distribution of ${\rho \propto r^{-\gamma}}$ transforms into a spike of slope ${\gamma_\text{GS} = (9-2\gamma)/(4-\gamma)}$. These spikes are predicted to start at one-fifth of the radius of gravitational influence of the central BH \citep{MerrittConferenceProceedings,Merritt}. While DM spikes have been of interest for the detection of DM, previous studies into their shape has predominately been (semi-)analytical. To the best of the authors' knowledge, fully numerical N-body simulations of the formation of these DM spikes have never been published. These allow for important checks of (semi-)analytical formalisms, and should lead to more realistic DM halo density profiles.

In this article, we present the first numerical step towards realistic spike density distribution functions, by demonstrating the formation of a DM spike in N-body simulations using realistic Hernquist halo profiles. In Section \ref{sec:theoreticalbackground}, we give the used DM halo distribution functions and analytical predictions of the expected spikes. In Section \ref{sec:numericalmethods}, we describe the used code and numerical schemes (Section \ref{sec:codeandnumericalschemes}), how the initial haloes are constructed (Section \ref{sec:initialisationofthehaloes}), and a description of the simulated systems and their processing (Section \ref{sec:simulatedsystemsandprocessing}). In Section \ref{sec:results}, we propose a new profile for the spike (Section \ref{sec:res:spikeprofile}), and contextualise the results by estimating their impact on the detectability of dephasing in GWs (Section \ref{sec:implicationsofresults}). Finally, in Section \ref{sec:conclusions}, we draw conclusions and some concluding remarks are given. Appendices with additional material regarding numerical solutions and validation of the simulations are also supplied after the bibliography.

\section{Theoretical Background}
\label{sec:theoreticalbackground}
Many different DM density profiles exist and are actively compared to observational data (e.g. \citealt{Li:2020iib}). Arguably the most prominent spherical profiles are the \citet*{NFW} (NFW) and the \citet{Hernquist} profiles. Both feature an inner cusp of $\rho \propto r^{-1}$, with the Hernquist having a more steep outer density falloff of $\rho \propto r^{-4}$ compared to $\rho \propto r^{-3}$ of the NFW.\footnote{See \citet{Springel_2005} for a more thorough discussion regarding their differences.} We will use the Hernquist profile, because it does not require the use of artificial cut-offs and is easier to work with analytically. Due to the identical inner cusp we expect the results presented in this article to be equivalent for both profiles. The density profile of a Hernquist halo is given by
\begin{equation}
    \label{eq:HernqDensity}
    \rho_\text{Hernq} = \frac{M_\text{h}}{2\pi}\frac{a}{r}\frac{1}{(r+a)^3},
\end{equation}
where $M_\text{h}$ is the total mass of the halo, and $a$ is the characteristic \textit{scale radius}. Its gravitational potential $\psi_\text{Hernq}$ is then given by
\begin{equation}
\label{eq:HernqPotential}
    \psi_\text{Hernq} = -\frac{G M_\text{h}}{r + a}.
\end{equation}
Simulations by \citet{Correa:relationConcentration} have shown that the two free parameters $a$ and $M_\text{h}$ are correlated for realistic Hernquist halos, and we thus define our haloes using only $M_\text{h}$. This article still includes $a$ in equations for ease of reading and consistency with other works. Assuming zero redshift, this relation becomes:
\begin{equation}
    \log_{10}\frac{r_\text{vir}}{a} = 1.4981 + 1.4540 \log_{10}\frac{M_\text{h}}{\text{M}_\odot}\left[1 + 0.0058 \log_{10}\left(\frac{M_\text{h}}{\text{M}_\odot}\right)^2\right],
\end{equation}
where $r_\text{vir}$ is the virial radius, defined as the radius where the encapsulated halo has a density of 200 times the critical density of the universe, $\rho_c = H^2/8\pi G^2$, with $H$ the Hubble parameter.\footnote{$\rho_c$ is the density required for a spatially flat universe in the FLRW metric. Note that in some literature, $r_\text{vir}/a$ is defined as the concentration $c$.}

In this study, we introduce a BH to the centre of Hernquist haloes, and grow it adiabatically. Specific growth rates have been proposed for BHs, such as the Bondi-Hoyle-Lyttleton accretion rate \citep{10.1093/mnras/101.4.227,10.1093/mnras/104.5.273}, however the final state of the system is independent of the actual rate of growth within the assumption of adiabatical growth, due to the invariance of the actions of the DM halo \citep{galacticDynamics}. Following G\&S, a spike of slope $\gamma_\text{sp} = 7/3$ is predicted for the inner regions of the halo. This spike is predicted to start at
\begin{equation}
\label{eq:gsspikerad}
    r_\text{sp, Hernq} = \alpha_\gamma a \left(2\pi\frac{M_\text{BH}}{M_\text{h}} \right)^{0.5} \,,
\end{equation}
for a Hernquist profile, where $\alpha_\gamma$ is a factor obtained from numerical integration, being $0.122$ for $\gamma = 1$. If the growth is non-adiabatic, a milder slope is predicted \citep{Ullio:2001fb}. The BHs in this study are treated as Newtonian point-masses, as relativistic effects on the haloes only become significant at radii of the order of 10 Schwarzschild radii $R_s$ \citep{Sadeghian:2013laa}. However, the exact shape of a G\&S spike close to the BH could be less steep based on the BH formation history \citep{RenskPaper}. This effect is again below our resolution, such that we predict our simulated spikes to follow a power-law. The growth of the BH would be fueled by a baryonic component present in the total astrophysical system. However, it can be excluded from the spike calculations as long as the total baryonic mass is much smaller than the mass of the combined BH + DM halo system, which is often the case in realistic systems and is also assumed by G\&S.

Numerical studies are often performed with either the NFW, Hernquist, or toy profiles of a constant slope of $\rho \propto r^{-1}$. A spiked DM halo profile is then often approximated as an initial profile with a spike added after a certain spike radius  $r_\text{sp}$ (e.g. \citealt{BertoneGammaRaysIMBH,Eda:2013gg,Eda:2014kra,Kavanagh:2020cfn,Mukherjee:2023lzn,gammaray}):
\begin{equation}
\label{eq:MerrittRadGravInfl}
    \rho(r) = 
\begin{cases}
    \rho_\text{init}(r=r_\text{sp})\left( \frac{r}{r_\text{sp}}\right)^{-7/3} &  r \leq r_\text{sp}\\
    \rho_\text{init}(r) & r > r_\text{sp}
\end{cases},
\end{equation}
where $r_\text{sp}$ is not taken to be the G\&S prediction, but instead as one-fifth of the radius of gravitational influence $r_h$, defined as the radius where the enclosed halo mass is equal to twice the mass of the central BH \citep{MerrittConferenceProceedings}:
\begin{align}
\label{eq:MerrittApprox}
4\pi\int_0^{r_h} \rho(r)r^2\der r &= 2M_\text{BH}.
\end{align}
When solved for the Hernquist halo, the radius of the spike becomes
\begin{equation}
\label{eq:MerritRadiusMu}
    r_\text{sp} = \frac{r_\text{h}}{5} = \frac{2a\mu + a\sqrt{2(\mu - \mu^2)}}{5(1-3\mu)},
\end{equation}
where $\mu = M_\text{BH}/M_\text{tot}$ is the mass ratio. We note that Equations \eqref{eq:MerritRadiusMu} and \eqref{eq:gsspikerad} are equivalent for $\mu \lessapprox 0.01$, but quickly deviate after.  We refer to the profile of G\&S, expanded using Equations \eqref{eq:MerrittRadGravInfl} and \eqref{eq:MerrittApprox}, as the \textit{Modified G\&S profile}. We note that this profile was originally derived for an isothermal density profile with $\rho \propto r^{-\gamma}$ for $0.5 \leq \gamma \leq 2$, and that deviations are thus expected as the Hernquist starts to deviate from the isothermal assumption.

All particles in our simulated haloes are gravitationally bound, and must thus have energies below the binding energy $\mathcal{E}$. The distribution function of these energies is the Eddington formula \citep{galacticDynamics}:
\begin{equation}
    f(\mathcal{E}) = \frac{1}{\sqrt{8} \pi^2M_\text{h}} \int_0^\mathcal{E} \frac{\partial^2 \rho}{\partial \psi^2} \frac{\der \psi}{\sqrt{\mathcal{E} - \psi}} + \frac{1}{\sqrt{8\mathcal{E}}\pi^2M_\text{h}} \left( \frac{\partial \rho}{ \partial \psi} \right)_{\psi = 0}.\label{eq:init:eddington2}
\end{equation} 
In appendix \ref{sec:appendix:eddingtonsolutions}, we give the solutions for both an isolated Hernquist halo, and one where a central BH is already present.

\section{Numerical Methods}
\label{sec:numericalmethods}
\subsection{Code and numerical schemes}
\label{sec:codeandnumericalschemes}
The simulations presented here are performed with a modified version of the \textsc{swift} code \citep{SWIFT:2023dix}.\footnote{\textsc{swift} is available on \url{www.swiftsim.com}, where one can also find extended documentation. Our modifications are based on version 0.9.0.} We introduced a new BH model to the code that can grow its mass over time at a uniform pace until a final mass is reached. This BH model is called the "\textbf{D}ark Matter \textbf{A}ttracting \textbf{B}lack Hole", or DAB for short. The BHs in this model have a set rate of growth and a final mass, after which the growth is stopped. Expanding DAB to grow following accretion formalisms where baryonic matter is deleted is easily done, and fully supported by \textsc{swift}, however we have chosen not to do so in order to reduce computational complexity. 

\textsc{swift}'s gravity solver employs the \textit{Wendland-C2} kernel, named after \citet{Wendland}, where the gravitational interactions between two particles are smoothed below a certain \textit{smoothing radius} $\epsilon$. This value cannot be taken too small, as to prevent two-body interactions, nor too large, since the resolution of our simulations is determined by this softening. We follow the scheme as developed by \citet{P03},
\begin{equation}
\label{eq:NBody:softening}
    \epsilon = \alpha \frac{r_\text{vir}}{\sqrt{N_\text{vir}}},
\end{equation}
where $\alpha$ is a numerical constant set empirically. While different values exist in the literature, we find $\alpha = 6$ to be stable in all our simulations.\footnote{E.g. \citet{softeningZhang} find a value of $\alpha = 2$, and \citet{P03} find a value of $\alpha = 4$, which cause numerical instabilities in some of our simulations.} The smallest radius yielding physical results is called the \textit{convergence radius} $r_\text{conv}$, the radius below which particles start to show two-body behavior \citep{duffy}. We found that $r_\text{conv} = 2\epsilon$ for our simulations after 4 Gyrs, where we refer to appendix \ref{sec:convergenceradius} for the full analysis. 

Having a large amount of particles in the simulation will thus lower $r_\text{conv}$ and improve our resolution. However, the complexity of the calculations performed by \textsc{swift} scale as $\mathcal{O}(N\log N)$ compared to the resolution scaling as $\mathcal{O}(1/\sqrt{N})$, and the balance between a manageable computation time and a small resolution is a fine one.\footnote{Appendix \ref{sec:appendix:influenceresolution} briefly touches upon this very quickly increasing complexity in the context of systemic error calculations.} As the mass of the central BH grows, and larger forces are to be computed, the timestepping of the simulation gets smaller. \textsc{swift} allows for dynamical timestepping $\Delta t$, where the timesteps are scaled to the forces acting upon the particle. The time-stepping is defined as \citep{P03}
\begin{equation}
    \Delta t_i = \sqrt{\frac{2\eta \epsilon}{\abs{\Vec{a_i}}}},
\end{equation}
where we found that $\eta = 0.005$ in our simulations. Due to this complexity and ever decreasing timestepping as the mass of the central BH grows, individual simulations took up to two weeks to finish. 

Finally, this article employs a numerical implementation of the G\&S formalism to compare our results to previous studies, allowing us to compute results for the exact same Hernquist haloes as the main N-Body simulations. We refer to this as the \textit{Numerical G\&S formalism}, and it is based on code developed for \citet{RenskPaper}.

\subsection{Initialisation of the haloes}
\label{sec:initialisationofthehaloes}
Every DM particle in our simulations has 7 properties: positions $(x,y,z)$, velocities $(v_x,v_y, v_z)$, and a mass set equal across all DM particles. The initial conditions are chosen so that an isolated halo is stable over the time span of the simulation, in the following way:
\begin{enumerate}
    \item $N$ DM particles are created and radially distributed using inverse transform sampling. The azimuthal and polar angles are sampled uniformly.
    \item A BH particle of desired mass is placed in the centre. In practice, this was set to be the DM mass as to not have a large impact on the distribution function.
    \item An array of binding energies $\mathcal{E}_x$ ranging from $0$ to $\psi$ -- Equation \eqref{eq:HernqPotential} -- is created for each DM particle. The distribution function of the energies in this array is numerically calculated,
    \begin{equation}
    F(\mathcal{E}_x) = \frac{\int_0^{\mathcal{E}_x} f(\mathcal{E})\sqrt{\psi - \mathcal{E}} \der\mathcal{E}}{\int_0^{\psi} f(\mathcal{E})\sqrt{\psi - \mathcal{E}} \der\mathcal{E}}.
\end{equation}
    where $f({\mathcal{E}})$ is the solution of the Eddington formula of Equation \eqref{eq:init:eddington2} (see also Appendix \ref{sec:appendix:eddingtonsolutions}).
    \item $\mathcal{E}_x$ is sampled from $F(\mathcal{E}_x)$, resulting in the velocity of the particle using $\abs{\Vec{v}} = \sqrt{2(\psi - {\mathcal{E}_x})}$. Its direction is taken at random.
\end{enumerate}
Using the above, all particles in the simulation are both bound and isotropically distributed as long as $N \gg 1$,\footnote{This is verified both before and during the simulations using the \textit{Shannon Entropy} of DM particles as viewed from the BH \citep{Entropy}.} but have otherwise completely free orbits. The central BH is not bound in place, and will move due to its gravitational interactions. This movement is completely random, but increases as the spike forms and the central densities around the BH increase. The halo is bound to the BH and it follows it around, making it so the final results are not significantly influenced by this movement.\footnote{For reference, a BH of mass $5\times 10^3 \text{M}_\odot$ inside a halo of mass $10^4 \text{M}_\odot$ moves $3.82 \times 10^{-2}$ kpc over 4 Gyr. Throughout this movement, the halo follows the BH perfectly such that it remains in the centre of its halo.}

These initial conditions produce a very small initial shock wave when the simulation is started with a BH present. This effect is thoroughly treated in Appendices \ref{sec:appendix:eddingtonsolutions} and \ref{sec:numericalshockwaves}, where we conclude that this of no significant effect to our final results.

\begin{table*}
    \centering
    \begin{tabular}{l|c|c|c|c|c|c|c}
    \hline
        Name & $M_\text{h}$ & $r_\text{vir}$ & $a$ & $c$ & $M_\text{BH}$ after 4 Gyr & Final $\mu$ & $N$ particles \\
        \hline 
        5e4-1e3 & $5\times 10^4 \text{M}_\odot$ & 0.78 kpc & 0.033 kpc & 23.2 & $10^3 \text{M}_\odot$ & 0.020 & $150^3$\\
        1e5-5e3 & $10^5 \text{M}_\odot$ & 0.98 kpc & 0.043 kpc & 22.7 & $5 \times 10^3 \text{M}_\odot$ & 0.048 & $150^3$\\
        1e4-1e3 & $10^4 \text{M}_\odot$ & 0.45 kpc & 0.019 kpc & 24.5 & $10^3 \text{M}_\odot$ & 0.074 & $130^3$\\
        5e4-5e3 & $5\times 10^4 \text{M}_\odot$ & 0.78 kpc & 0.033 kpc & 23.2 & $5 \times 10^3 \text{M}_\odot$ & 0.091 & $130^3$\\
        5e3-1e3 & $5\times 10^3 \text{M}_\odot$ & 0.36 kpc & 0.014 kpc & 25.0 & $10^3 \text{M}_\odot$ & 0.167 & $130^3$\\
        3e3-1e3 & $3\times 10^3 \text{M}_\odot$ & 0.30 kpc & 0.012 kpc & 25.4 & $10^3 \text{M}_\odot$ & 0.250 & $130^3$ \\
        1e4-5e3 & $10^4 \text{M}_\odot$ & 0.45 kpc & 0.019 kpc & 24.5 & $5\times10^3 \text{M}_\odot$ & 0.333 & $130^3$\\
        \hline
    \end{tabular}
    \caption{Overview of the different runs of presented in this article. The naming scheme follows the format "$M_\text{h}$-$M_\text{BH}$", where $M_\text{BH}$ is the BH mass after 4 gyrs.}
    \label{tab:differentruns}
\end{table*}

\subsection{Simulated systems and processing}
\label{sec:simulatedsystemsandprocessing}
Simulations and detections show that Super Massive BHs in the centres of galaxies are orders of magnitude lighter compared to their surrounding haloes (e.g. \citealt{Ferrarese:2002ct,Bandara_2009,Booth:2009zb}). However, simulating these kinds of systems would place the spike fully within $r_\text{conv}$ and thus unresolvable. We furthermore use IMBHs due to the computational difficulty of a system with a SMBH and halo of the same order of magnitude. Therefore, this paper simulates systems where the final BH mass is between $10^3 \text{M}_\odot$ and $5\times10^3 \text{M}_\odot$, and the mass of the corresponding halo is of equal order of magnitude or one higher. As a result, the mass ratios $\text{M}_\text{BH}/\text{M}_\text{h}$ in our study are higher than those treated by previous studies (e.g. \citealt{Bandara_2009,Booth:2009zb}).

We deem that these systems will yield results that can also be applied to larger haloes, as long as the mass ratio stays the same. For $r\ll a$, the density slope is equal, and the density $\rho \approx M_\text{h}/2\pi a^3\tilde{r}$, with $\tilde{r} = r/a$. The radial velocity dispersion $\bar{v_r^2}$ is also equivalent in the same regime up to a scaling factor with $\bar{v_r^2} \approx GM_\text{h}/\tilde{r}\ln({1/\tilde{r}})$ \citep{Hernquist}. The internal dynamics of larger and smaller haloes are thus equivalent up to scaling factors.

We validate that the rate of growth of the central BH is adiabatic by growing a $10^3$$\text{M}_\odot$ BH using different rates-of-growth in a $10^4\text{M}_\odot$ halo, varying from 250$\text{M}_\odot$/Gyr to 4000$\text{M}_\odot$/Gyr. The resulting spikes are indistinguishable above the convergence radius for rates of growth of 2000$\text{M}_\odot$/Gyr and below. The highest absolute and relative rate-of-growths of the simulations are below this reference value. We also compute the orbital times of particles in a Hernquist halo in Appendix \ref{sec:dynamicaladiabatic}, and verify that those within the G\&S spike radius have orbital times below the maximum simulation time of 4 Gyr. We thus conclude that the assumption of adiabatic growth is valid.

\begin{figure}
    \centering
    \includegraphics[width=\linewidth]{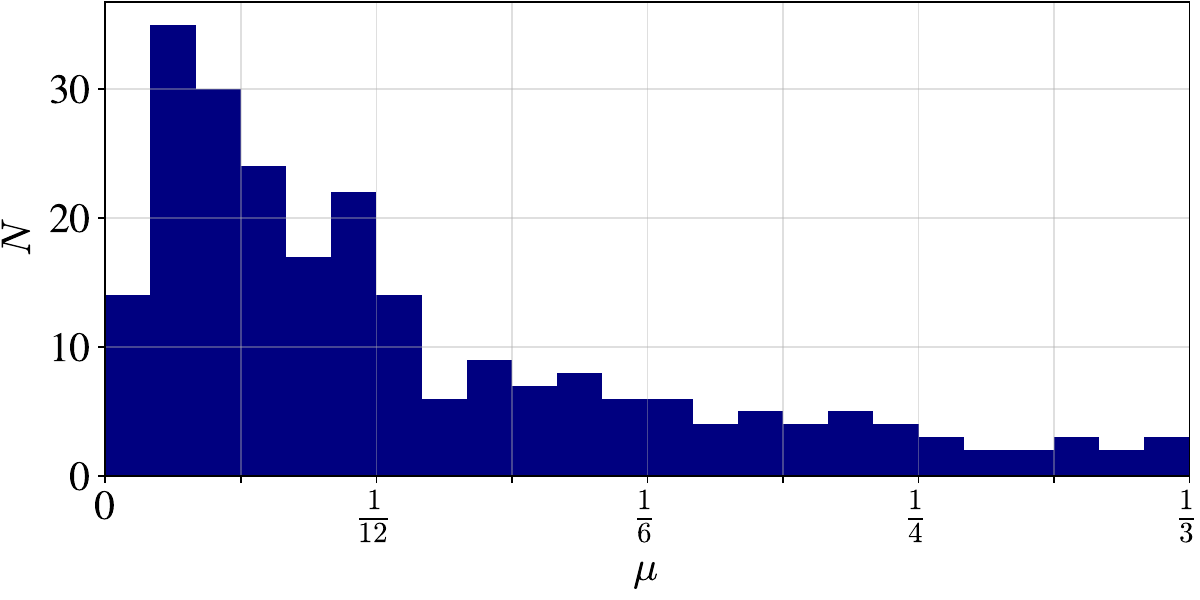}
    \caption{ The distribution of mass ratio $\mu$ in the present simulations, with $\mu = M_\text{BH}/M_\text{tot}$. Only systems where a spike is measured are included, for a total of 235 datapoints. As all BHs are grown from near-zero mass, lower values of $\mu$ are overrepresented.}
    \label{fig:histogram}
\end{figure}

The simulations were performed in 7 different runs, each with a different halo mass and final BH mass, grown over a period of 4 Gyr. Every 0.1 Gyr, the system is recorded, and as the growth is adiabatic, each of these snapshots is a result. This creates a large parameter space of 235 different values of $\mu$, see also Figure \ref{fig:histogram}.  The simulated systems were chosen such that the largest possible range of $\mu$ is probed, and are given in Table \ref{tab:differentruns}. Some lower mass ratios did not produce measurable spikes, as these manifest below $r_\text{conv}$. Due to the BHs growing from near-zero mass, low values of $\mu$ are overrepresented in the final dataset.

The final particle locations are radially binned from the location of the BH, and are assumed to be Poisson-distributed in said bins. The error of such a distribution are given by $\sqrt{N_\text{bin}}$, and as more than $10^4$ particles are present in the least populated bins, these errors are relatively very small. These histograms are then fitted with a least-squares procedure using Minuit through the PyRoot package \citep{minuit,ROOT}. The lower fitting boundary was taken to be the radius of convergence $r_\text{conv}$, and the upper was chosen such that the fitting stops below the small numerical shock waves discussed above.

\section{Results}
\label{sec:results}
Our simulations produce measurable DM spikes. An example of these spikes can be found in Figure \ref{fig:examplefit}, where the the normalised density is fitted as a function of radius. Also visible is a numerically calculated G\&S spike, which shows a clear deviation. In total, 235 unique systems with spikes were recorded, demonstrating the existence of this phenomenon in numerical simulations. This large dataset allows us to fit a DM spike profile, where normalised parameters were found to be only dependent on the mass ratio $\mu$. We first discuss our proposed profile, compare it to used approximations in the field, and then make an estimation of the effect of these results on the dephasing GWs due to DM spikes. 

\subsection{Spike Profile}
\label{sec:res:spikeprofile}
We empirically propose the final profile of a Hernquist halo with a spike after central adiabatic BH growth to be
\begin{align}
    \label{eq:spikeProfile}
    \rho_\text{final} &= \frac{M_\text{h}}{2\pi}\frac{a}{r}\frac{1}{(r+a)^3}\left[\beta + \left(\frac{r}{a} \frac{1}{\Tilde{r}_\text{sp}}\right)^{1 - \gamma_\text{sp}}\right],\\
    \label{eq:spikeProfileNormalised}
    \tilde{\rho}_\text{final} &= \beta + \left(\frac{r}{r_\text{sp}}\right)^{1 - \gamma_\text{sp}}
\end{align}
where $\tilde{\rho}_\text{final}$ is the final density normalised by $\rho_\text{Hernq}$, $\tilde{r}$ is radius normalised by $a$, $r_\text{sp}$ is the spike radius, $\beta$ governs the depletion of the original Hernquist profile as particles are pulled inward, and $\gamma_\text{sp}$ is the slope of the spike. This profile is equivalent to (modified) G\&S when $\beta$ is set to one.

\begin{figure*}
    \centering
    \begin{subfigure}{0.49\linewidth}
        \centering
    \includegraphics[width=\linewidth]{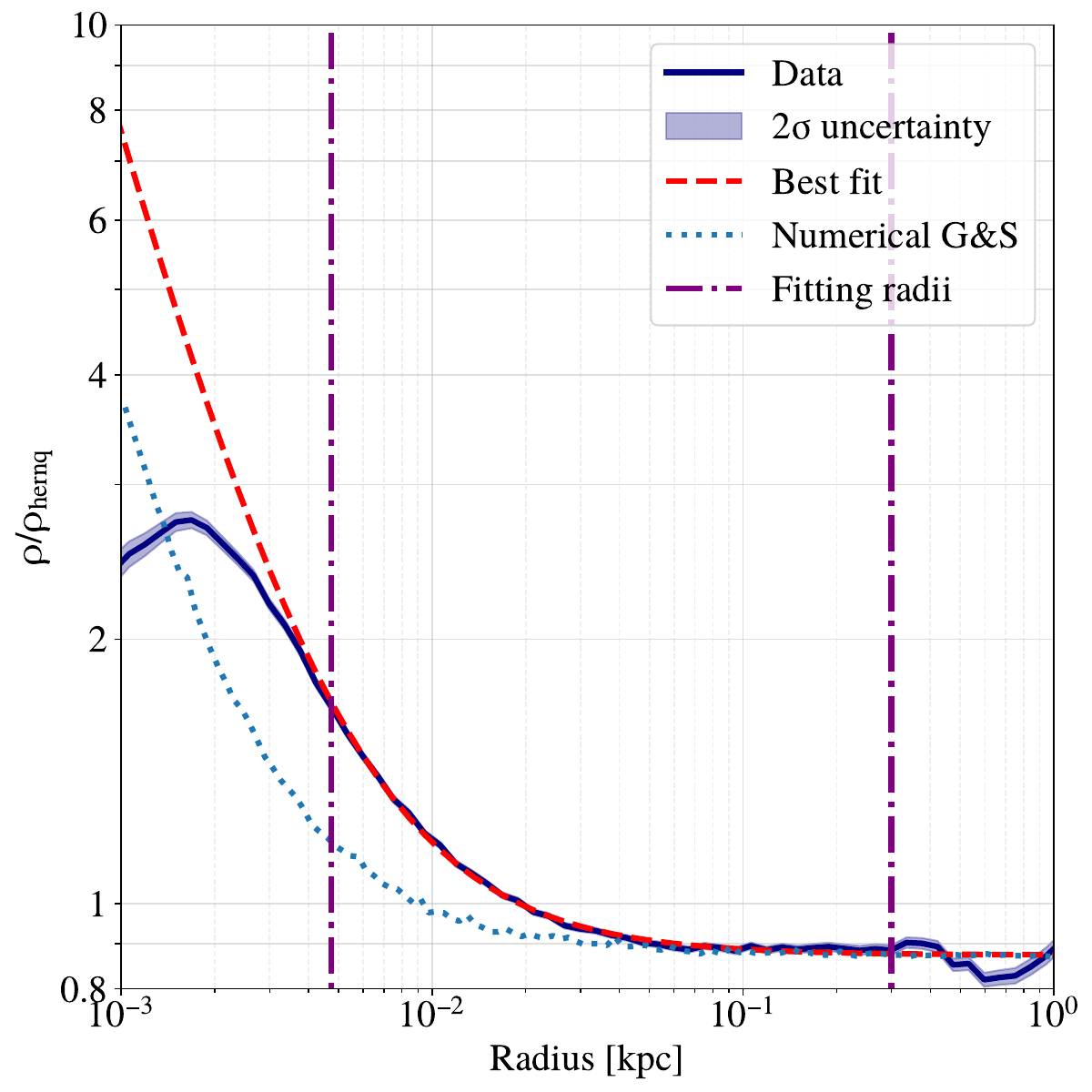}
    \caption{}
    \label{fig:examplefit}
    \end{subfigure}
    \begin{subfigure}{0.49\linewidth}
        \includegraphics[width=\linewidth]{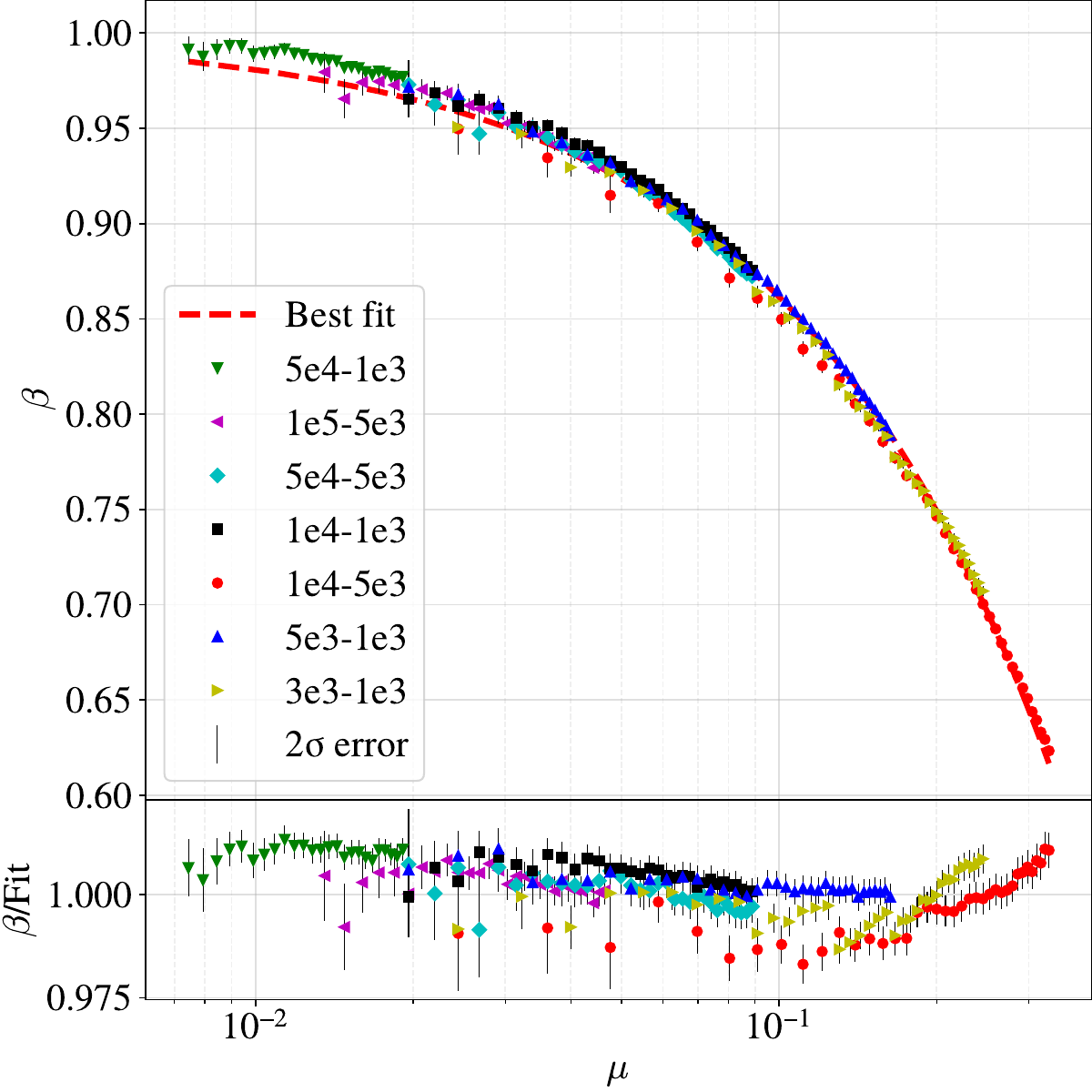}
    \caption{}
    \label{fig:fitbeta}
    \end{subfigure}
    \begin{subfigure}{0.49\linewidth}
        \includegraphics[width=\linewidth]{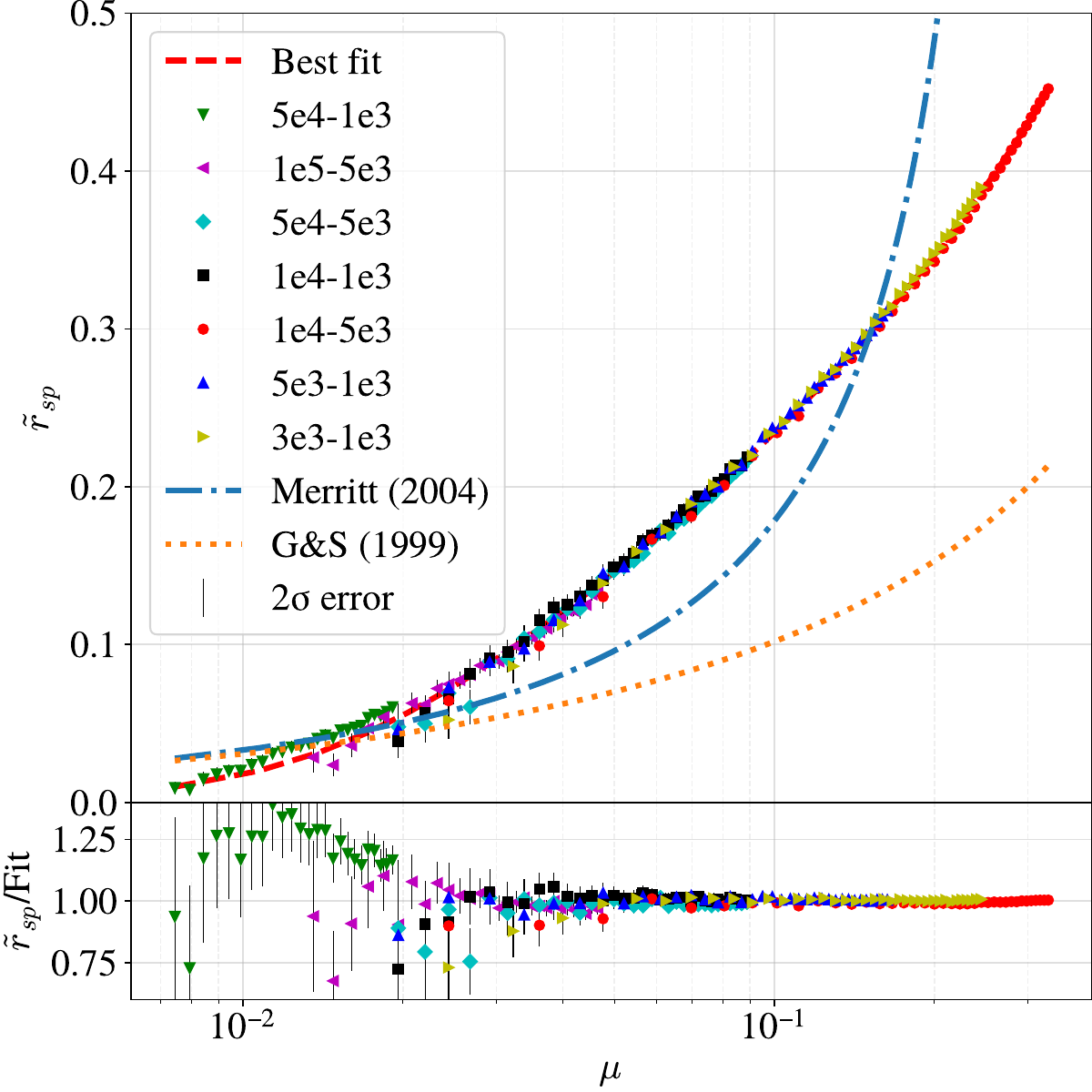}
    \caption{}
    \label{fig:fitrsp}
    \end{subfigure}
    \begin{subfigure}{0.49\linewidth}
        \centering
    \includegraphics[width=\linewidth]{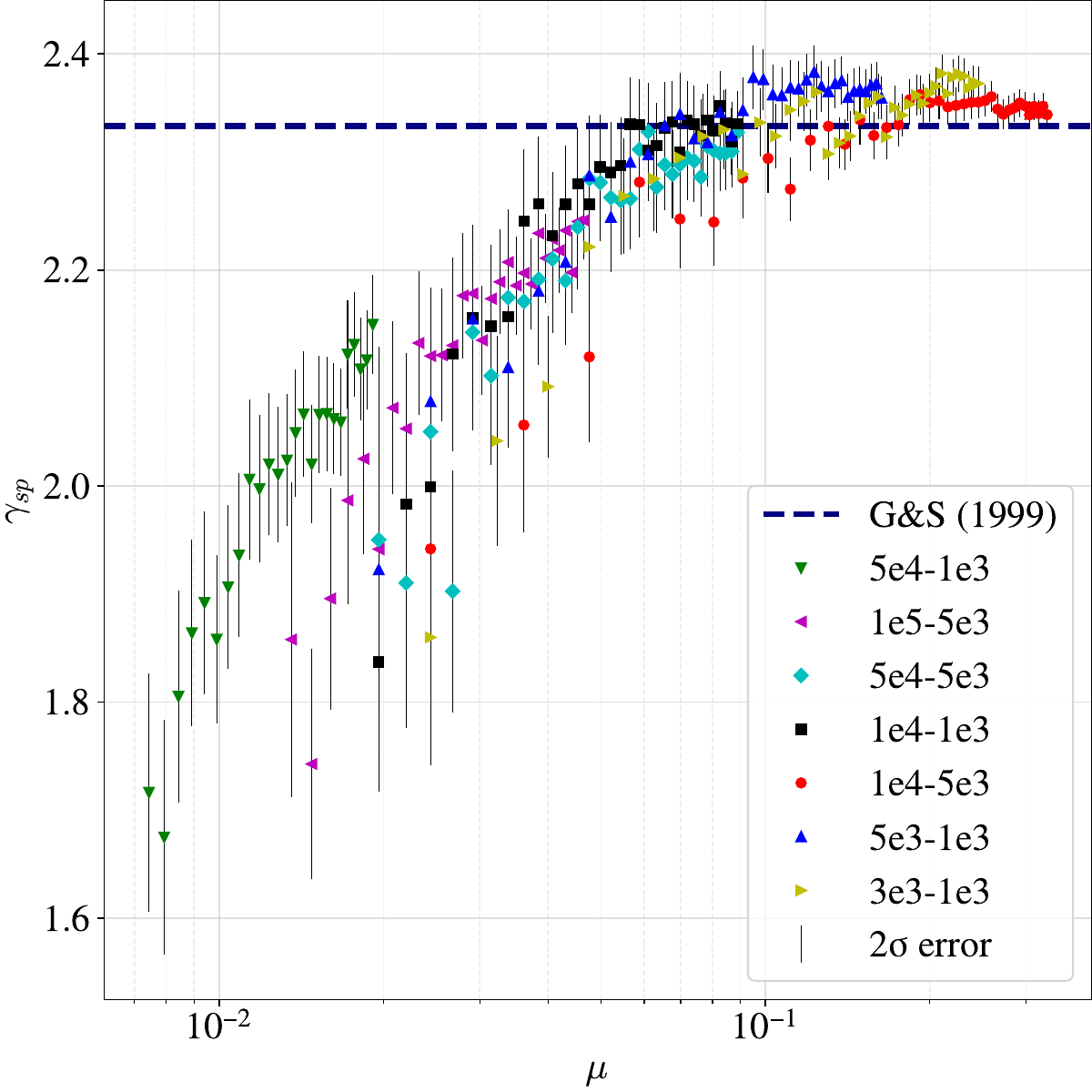}
    \caption{}
    \label{fig:slope_param}
    \end{subfigure}
    \caption{The fitted values of every parameter of the spike profile $\rho = \rho_\text{Hernq}[\beta + (r/r_\text{sp})^{1 - \gamma_\text{sp}}]$, shown with $2\sigma$ error. The fitted systems are found in Table \ref{tab:differentruns}. \textbf{Top left:} The simulated spike for a halo of $M_\text{h} = 10^4 \text{M}_\odot$ and a BH of mass $M_\text{BH} = 10^3 \text{M}_\odot$. The solid blue line shows the simulated data, with the shaded area being the 2$\sigma$ uncertainty. The dashed red line shows the best fit, being fitted between the two purple dash-dotted lines (see Appendix \ref{sec:convergenceradius}). Finally, the light blue dotted line shows the numerically calculated result from \citet{Gondolo} using \citet{RenskPaper}. \textbf{Top right:} The best fits for the halo depletion parameter $\beta$ as a function of $\mu$. The red dashed line is the best fit using Equation \ref{eq:beta}. The lower part shows the values of $\beta$ divided by the best fit. \textbf{Bottom left:} The best fits for the spike radius $r_\text{sp}$ as a function of $\mu$. The red dashed line is the best fit using Equation \ref{eq:fitrsp}. The lower part shows the values of $\beta$ divided by the best fit. \textbf{Bottom right:} The best fits for the spike slope $\gamma_\text{sp}$ as a function of $\mu$. The dashed line in blue is the value of 7/3 as found by \citet{Gondolo}.}
\end{figure*}

We check whether this profile can be fitted within our limited fitting range in Appendix \ref{sec:appendix:influenceresolution}, where we fit artificial data with Gaussian noise, both over a large and a limited range. While the fitting uncertainties do increase as the fitting range decreases, we observe no systematic differences between the two fits. The fitting uncertainties for $\beta$ and $r_\text{sp}$ are stable within an order of magnitude, but the uncertainty on $\gamma_\text{sp}$ grows significantly. We thus expect the fits of the N-body data to yield precise values for $\beta$ and $r_\text{sp}$, while $\gamma_\text{sp}$ will be harder to determine.

We present the fits of all systems in Figures \ref{fig:fitbeta}, \ref{fig:fitrsp} \& \ref{fig:slope_param} as a function of the mass-ratio $\mu$. Both $\beta$ and $r_\text{sp}$ show a clear dependence on $\mu$, and we thus perform further fits to determine their functional forms. The halo depletion parameter $\beta$ is fitted to a simple power law
\begin{equation}
\label{eq:beta}
    \beta = 1  - \alpha_1 \mu^{\alpha_2} \quad \quad
    \begin{cases}
        \alpha_1 &= 0.998 \pm 0.008\\
        \alpha_2 &= 0.858 \pm 0.004\\
    \end{cases},
\end{equation}
with the errors given in 1$\sigma$, and a goodness-of-fit of ${\chi^2_\text{red} = 1.81\times 10^{-5}}$. The data-points and resulting fit are given in Figure \ref{fig:fitbeta}. For $\mu = 0$, $\beta$ is imposed to be 1. We note that for small $\mu$ the relative depletion is approximately linear, and independent of the halo mass.

The spike radius $\tilde{r}_{\text{sp}}$ is fitted to a more complex function
\begin{equation}
\label{eq:fitrsp}
    \tilde{r}_\text{sp} = \alpha_3 \frac{\mu^{\alpha_4}}{\mu^{\alpha_5} + \alpha_6} \quad \quad
    \begin{cases}
        \alpha_3 & 0.801 \pm 0.004\\
        \alpha_4 & 2.29 \pm 0.07 \\
        \alpha_5 & 1.78 \pm 0.07 \\
        \alpha_6 & (9.1 \pm 2.4) \times 10^{-4}
    \end{cases},
\end{equation}
with the errors given in 1$\sigma$, and a goodness-of-fit of ${\chi^2_\text{red} = 3.36 \times 10^{-6}}$. For systems where $\mu \gtrapprox 0.1$, this relation simplifies to $\tilde{r}_\text{sp} = 0.8\sqrt{\mu}$. The data-points and resulting fit are given in Figure \ref{fig:fitrsp}. Also included are the spike radius as found by G\&S (Equation \ref{eq:gsspikerad}) and the Modified G\&S profile (Equation \ref{eq:MerritRadiusMu}). The root mean squared error (RMSE) between every profile and the data is calculated and given in Table \ref{table:RMSE}. The RMSE is lowest for our profile by two orders of magnitude, showing a very clear improvement over previous models for the spike radius.

The disagreement between the data and the theoretical profiles appears both at high values of $\mu$ and at low values of $\mu$. At high values of $\mu$, we find $\tilde{r}_\text{sp}$ to be larger than predicted by G\&S. We suspect that this is due to the (lack of) self-consistency in the G\&S formalism compared to the original formalism derived by \citet{Young_1980}, which included multiple iterations to allow for changes in the density distribution (and thus the gravitational potential) of the stellar distribution during adiabatic growth. We suspect that this feedback effect is largest at large values of $\mu$, as the pull of the DM halo is significant. At low values of $\mu$, the data shows smaller spike radii than predicted by G\&S and Modified G\&S. However, we refrain from hypothesising about the cause as this could be the result of low resolutions of these small spikes.
\begin{table}
\begin{tabular}{l|l|l|l}
      \hline &Best Fit $r_\text{sp}$ & \citet{Gondolo} & Modified G\&S  \\
     \hline RMSE: & $4.4\times10^{-3}$ & $0.12$  & $1.2$ \\
     \hline
\end{tabular}
\caption{The RMSE as calculated for the Best Fit of $r_\text{sp}$, the spike radius as given by G\&S, and by the Modified G\&S profile of Equation \ref{eq:MerritRadiusMu}. The RMSE is given by $\sqrt{\sum_i^N (y_{i} - f(x_i))^2/N}$.}
\label{table:RMSE}
\end{table}

Our simulations did not find a singular value for the spike slope as predicted, with the values for large $\mu$ agreeing with the G\&S value of $\gamma_\text{sp} = 7/3$. For lower values of $\mu$, this spike slope becomes smaller. We suspect that this is due to the relatively large fitting uncertainty (as discussed in Appendix \ref{sec:appendix:influenceresolution}). The data points are given in Figure~\ref{fig:slope_param}. As $\mu$ increases, the error of the data decreases and their values converges to the value as predicted by G\&S. This is most likely due to the spike growing, and as a result being more visible in the data, resulting in a better fit. Concluding, for $\mu\gtrapprox0.06$, we find that $\gamma_\text{sp}$ is in agreement with the value found by G\&S. For values below this, we find lower values, likely due to limited fitting range as discussed in Appendix \ref{sec:appendix:influenceresolution}. More work needs to be done to show whether the low $\mu$ values will also converge to $\gamma_\text{sp} = 7/3$, or if the true value of $\gamma_\text{sp}$ is indeed smaller.

Substituting Equations \ref{eq:beta} \& \ref{eq:fitrsp} into our spike profile (Equation \ref{eq:spikeProfile}) and integrating this profile over space for $\gamma_\text{sp} = -7/3$ demonstrates mass conservation with deviations of less than $5\%$ for values of $\mu \leq 0.05$, and less than 10\% for values of $\mu \leq 0.1$.

\subsection{Implications for future detections}
\label{sec:implicationsofresults}

We demonstrate the impact of our proposed spike profile (Eq. \eqref{eq:spikeProfile}) compared to currently used approximations by simulating BH inspirals for both the proposed profile and Modified G\&S. The feedback on the inspiralling object differs between different halo profiles, generating different dephasings in the resulting GWs.  This has been done using the \textsc{HaloFeedback} code \citep{Kavanagh:2020cfn}.\footnote{The authors are aware of the more recent codes such as those presented by \citet{Mukherjee:2023lzn} and \citet{Kavanagh:2024lgq}, yielding amplified results for the dephasing. These codes were however not yet publicly available or are computationally very costly. As we only want to give an indication of the difference between the two profiles, this code suffices.} As $\gamma_\text{sp}$ is not uniquely determined, we compute inspirals in our proposed profile for both $\gamma_\text{sp} = 2$ and $7/3$.

We can quantify the dephasing by calculating the number of cycles that the secondary object travels between two times as
\begin{equation}
    \label{eq:Ncycles}
    \Delta N_\mathrm{cycles} (t_i, t_f) = \int^{t_f}_{t_i} \der t \; f_\mathrm{GW}(t) \,,
\end{equation}
where we fix $t_f = 0$ as the time that the secondary object reaches $r = 4 R_S$. We then fix $t_i$ as five years before $t_f$, regardless of environment. This allows us to make a comparison between a vacuum inspiral and an embedded inspiral. 

\begin{table}
    \centering
    \begin{tabular}{l l l l}
        \hline $M_\text{h}$ & Mod. G\&S & Eq. \eqref{eq:spikeProfile}, $\gamma_\text{sp} = 7/3$ & Eq. \eqref{eq:spikeProfile}, $\gamma_\text{sp} = 2$ \\
        \hline\noalign{\smallskip}
        $10^4 \text{M}_\odot$ & $910$ & $1900$ & $1.6$  \\
        $10^5 \text{M}_\odot$ & $560$ & $350$ & $0.6$\\
        \noalign{\smallskip}\hline  
    \end{tabular}
    \caption{The difference in number of cycles $\Delta N_\text{cycles}$ between a vacuum inspiral and an embedded inspiral for six different spikes, calculated using Eq. \eqref{eq:Ncycles}. The primary black hole has $M_\text{1} = 10^3 \text{M}_\odot$, while the secondary object has $M_\text{2} = \text{ M}_\odot$. The vacuum inspiral is of 3175960 cycles. If $\gamma_\text{sp}$ is indeed $-7/3$, then the dephasing in our profile is roughly similar to the Modified G\&S profile. For $\gamma_\text{sp} = -2$, the dephasing has almost disappeared.}
    \label{tab:Ncycles}
\end{table}

In Table \ref{tab:Ncycles}, the difference in the number of cycles $\Delta N_\text{cycles}$ is given for a toy system with a central BH $M_\text{1} = 10^3 \text{M}_\odot$ and an inspiralling object of $M_\text{2} = \text{M}_\odot$. For $M_\text{h} = 10^4 \text{M}_\odot$ we see an increase of $\Delta N_\text{cycles}$ of a factor 2 when our proposed profile is used compared to Modified G\&S, with a $\gamma_\text{sp} = 7/3$. However, when the slope is dulled to $\gamma_\text{sp} = 2$, $\Delta N_\text{cycles}$ decreases until it is practically naught. We note that for the corresponding value of $\mu$, our simulation found $\gamma_\text{sp} \approx 7/3$. 

If the halo mass is increased to $M_\text{h} = 10^5 \text{M}_\odot$, our profile yields slightly lower results of about thirty percent compared to Modified G\&S for $\gamma_\text{sp} = 7/3$. This is caused by the slightly smaller value of $r_\text{sp}$ at low $\mu$, which in turn lowers the normalisation of the spike density. This is also the $\mu$ regime where our simulations indicate a potentially lower value for the slope: $\gamma_\text{sp} = 2$ again shows $\Delta N_\text{cycles}$ nearly disappear. More shallow spikes thus yield a significant decrease of the dephasing of GWs compared to often used approximations. If $\gamma_\text{sp}$ is instead $7/3$ as is predicted by theory, then the dephasing effect is decreased by thirty percent, a smaller yet still significant effect.

\section{Conclusions}
\label{sec:conclusions}

The mass distribution of DM around BHs is of vital importance to some astrophysical DM detection methods. We demonstrated the formation of DM spikes in fully numerical N-body simulations using realistic Hernquist halo profiles. The constant inner slope of $\rho \propto r^{-1}$ means our results should be equivalent for other profiles such as the NFW, especially at low mass ratio $\mu$. We proposed a new profile for these DM spikes, which deviates from theoretical predictions and common approximations. This profile only depends on a single new parameter compared to the initial halo profile, the mass ratio of BH and the total system $\mu$. We observe a depletion of the original Hernquist of up to 20\%, and show the spike radius follows a different dependence on $\mu$. This impacts the overall normalisation of the spike compared to previous assumptions. We found a value of $\gamma_\text{sp} = 7/3$ for large values of $\mu$, but were unable to confidently determine it for smaller mass ratios due to the nature of the simulations. The impact of these results on future DM detections has been demonstrated using the expected dephasing of gravitational waves due to the presence of a DM halo. This dephasing was shown to deviate between our profile and previous approximations, increasingly so as the spike slope and mass ratio $\mu$ increases.

Our study unlocks the potential for future investigations into the exact shape of DM spikes and predictions for future DM detections. An improved resolution would yield smaller errors on the fitted parameters of the proposed profile, especially the slope, and allow for lower $\mu$ systems to be probed. Simply increasing the number of particles is not feasible however, due to the scaling nature of N-body simulations. If achieved however, an exact value for the spike slope can be determined and larger mass differences in the system can be probed, bringing us closer to realistic DM spike profiles. 

\section*{Acknowledgements}

JLK offers their sincerest gratitude to Camila Correa for the plentiful and insightful discussions on the code that was developed for this work. The insightful comments of the anonymous peer reviewer were instrumental to solving numerical issues and improving the quality of the manuscript, and both authors wish to express their sincerest gratitude for them. The authors furthermore express their sincerest gratitude to Tim Linden for insightful feedback on an early draft of this manuscript, and to Gerben Wierda for crucial technological support. JLK acknowledges SURF and the Snellius supercomputer on which part of the research was conducted. The authors acknowledge the work of the \textsc{swift}- and other collaborations, providing us with open-source software that is vital to scientific research, specifically the \textsc{MatPlotLib}, \textsc{Numpy} and \textsc{SciPy} python packages.

JLK performed the N-Body simulations underlying this article, and the resulting data was jointly processed by both authors. JLK analysed the results with critical input from RW. RW performed the analysis of the dynamical time-scales of the systems, the cross-reference of the data using the numerical G\&S formalism, and the inspiral analysis. JLK wrote the manuscript with critical input from RW. No AI was used during any part of the research underlying, and writing of, this work.

\section*{Data Availability}

The code underlying this article is a modified version of \textsc{swift}, and can be found at \url{https://github.com/JLKamermans/SWIFT-DAB}. The data is reproducible using this code, and will be shared upon reasonable request to the corresponding author.


\bibliographystyle{mnras_arxiv.bst}
\bibliography{bibliography}




\appendix

\section{Solutions to the Eddington Equation and initial conditions}
\label{sec:appendix:eddingtonsolutions}
The solution to the Eddington Equation for an isolated Hernquist halo is given by \citep{galacticDynamics}
\begin{equation}
    f(\mathcal{E}) = \frac{3 \arcsin{q} + q\sqrt{1 - q^2}}{\sqrt{2} (2\pi)^3(GM_\text{h}a)^{3/2}}\frac{(1 - 2q^2)(8q^4 - 8q^2 - 3)}{(1-q^2)^{5/2}}, \label{eq:init:eddingtonCDMnoBH}
\end{equation}
where  $q = \sqrt{a\mathcal{E}/GM_\text{h}}$. It is also possible to determine $f(\mathcal{E})$ with a massive BH already present. After the extra term $-GM_\text{BH}/r$ is added to $\psi$, the analytical solution involves a collection of elliptic integrals \citep{dynamicalStructure}, and a more manageable numerical result is preferred. Calculating the double derivative of $\rho$ w.r.t. $\psi$ yields
\begin{align}
 \frac{d^2 \rho}{d \psi^2} &= \frac{d}{d\psi}\left(\frac{d\rho}{dr} \frac{dr}{d\psi}\right) = \frac{d}{dr}\left( \frac{d\rho}{dr} \left(\frac{d\psi}{dr} \right)^{-1} \right) \left(\frac{d\psi}{dr} \right)^{-1}\\
 &=  \frac{ar^3}{G^2M_\text{tot}^2 \pi (a + r)} \frac{a^2 (1 + 5 \mu) + 6r^2 + 4a (r + 2 \mu r) }{(a^2 \mu + 2 a \mu r +r^2)^3}.
 \label{eq:init:d2nudpsi2}
\end{align}
Rewriting $r$ in terms of $\psi$ yields
\begin{equation}
    r(\psi) = - \frac{GM_\text{tot} + a\psi + \sqrt{- 4 G a M_\text{BH} \psi + (M_\text{tot} + a\psi)^2}}{2\psi}.
    \label{eq:init:rofpsi}
\end{equation}
Equation \eqref{eq:init:rofpsi} is to be substituted into Equation \eqref{eq:init:d2nudpsi2}, which is then inserted into Equation \eqref{eq:init:eddington2} for numerical integration. In the limit with no BH, $\mu = 0$, and the result of Equation \eqref{eq:init:eddingtonCDMnoBH} is obtained. The haloes generated using these initial conditions are found to be largely stable over a time span of 4 Gyr, with only a small numerical shockwave present that quickly travels radially outward. This shockwave is discussed further in Appendix \ref{sec:numericalshockwaves}. In figure \ref{fig:isolatedHalo}, two Hernquist haloes are shown after 4 Gyr: one without a central BH, and one with a central BH of constant mass equal to the initial mass as utilised in this work. Both show full equivalence to the initial Hernquist profile until the radius of convergence, below which a core forms due to numerical effects. The isolated Hernquist halo has also been simulated using the unmodified \textsc{swift} code upon which our modifications were built and an identical result was obtained, which is also included in figure \ref{fig:isolatedHalo}.

\begin{figure}
    \centering
    \includegraphics[width=\linewidth]{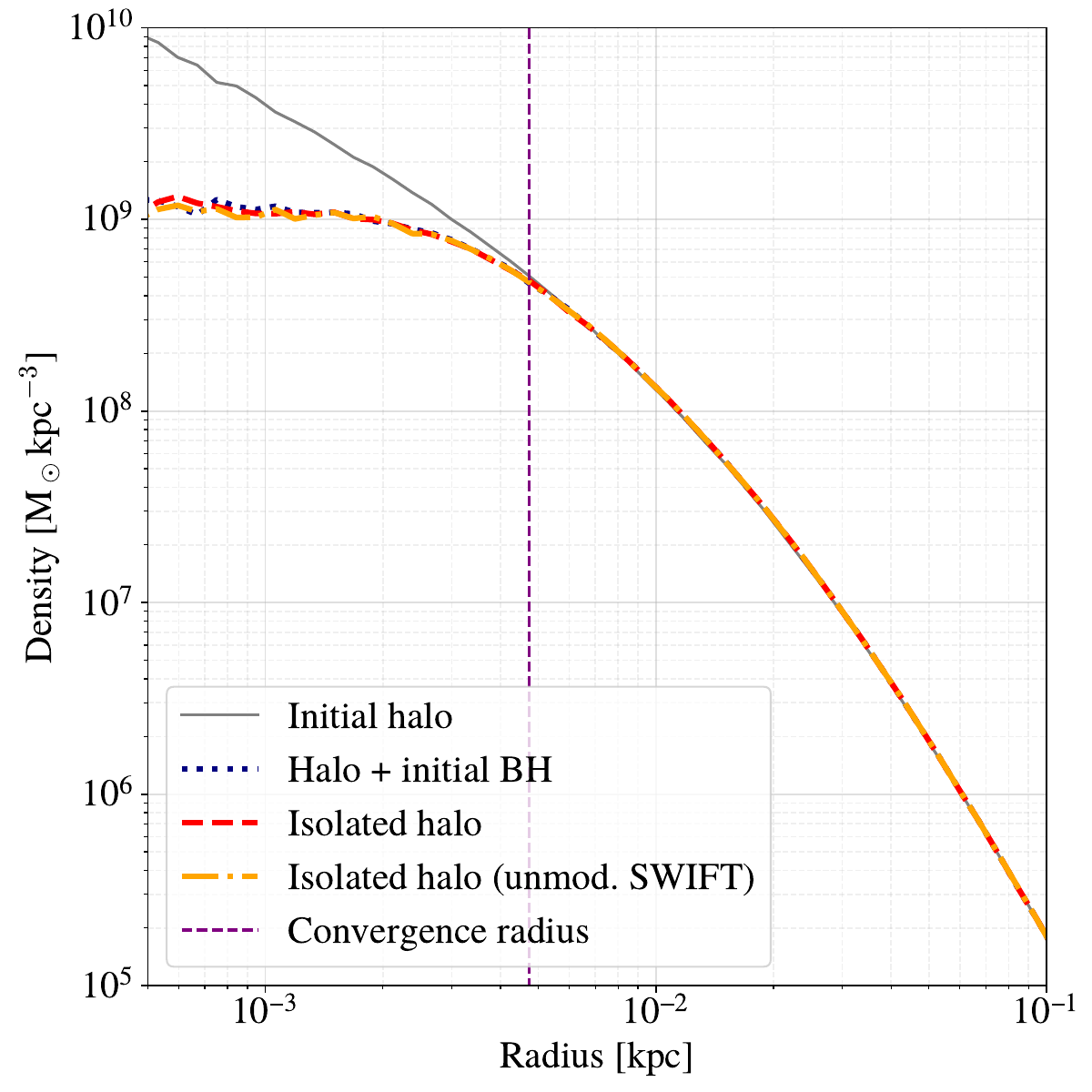}
    \caption{Two $10^4$ $\text{M}_\odot$ Hernquist haloes consisting of $130^3$ DM particles, simulated for 4 Gyr using the initial conditions of Appendix \ref{sec:appendix:eddingtonsolutions}. The grey straight line is the initial halo, the red dashed line is an isolated halo consisting of only DM, and the blue dotted line is the same halo with a central BH of the initial mass as utilised in this work ($4.55\times10^{-3}\text{M}_\odot$, the same mass as DM particles). Note that the BH does not grow in mass. The isolated halo is also simulated using the unmodified version of \textsc{swift}, shown in the orange dash-dotted line. The convergence radius is shown in the thin vertical purple dashed line.}
    \label{fig:isolatedHalo}
\end{figure}

\section{Determination of the Convergence Radius}
\label{sec:convergenceradius}
\begin{figure}
    \centering
    \includegraphics[width=\linewidth]{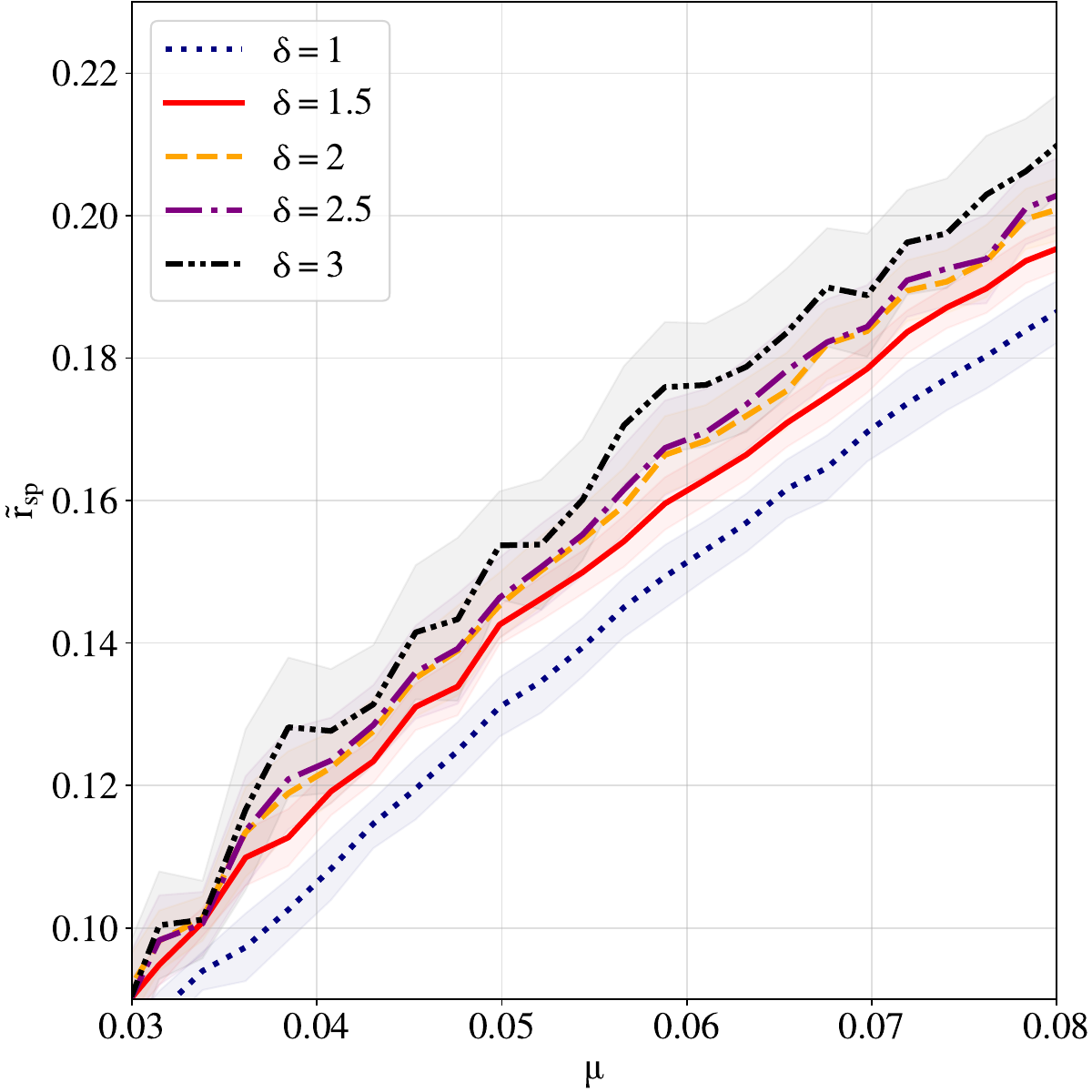}
    \caption{Fitted values of spike radius $r_\text{sp}$ of Equation \eqref{eq:spikeProfile} for a $10^4 \text{M}_\odot$ halo and a $10^3 \text{M}_\odot$ BH grown over 4 Gyr. Every line represents a different value of $r_\text{conv}$ during fitting, where $r_\text{conv} = \delta \epsilon$, and $\epsilon$ is the gravitational softening of Equation \eqref{eq:NBody:softening}. The plot has been zoomed in for visual clarity.}
    \label{fig:rconv1}
\end{figure}

\begin{figure}
    \centering
    \includegraphics[width=\linewidth]{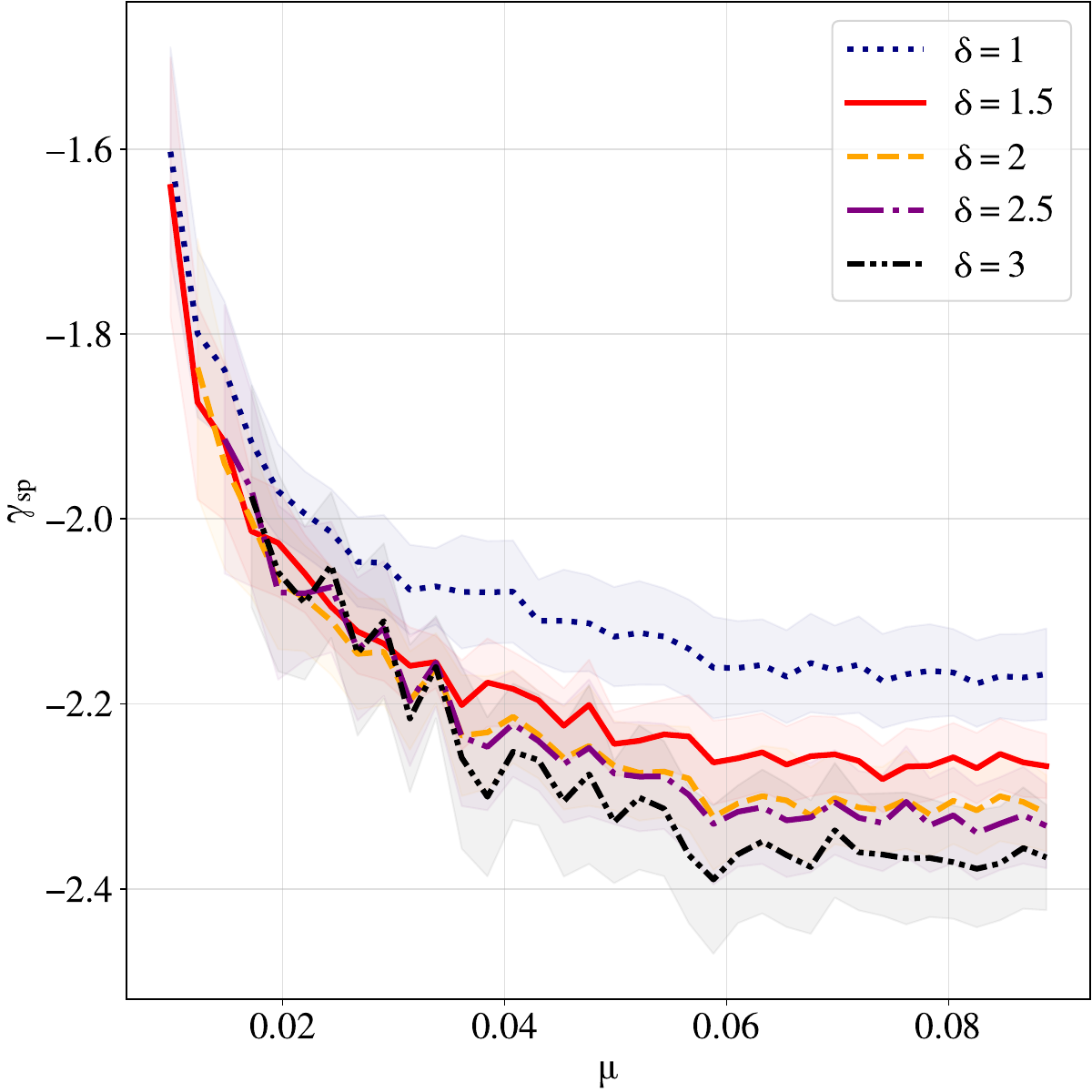}
    \caption{Fitted values of spike slope $\gamma_\text{sp}$ of Equation \eqref{eq:spikeProfile} for a $10^4 \text{M}_\odot$ halo and a $10^3 \text{M}_\odot$ BH grown over 4 Gyr. Every line represents a different value of $r_\text{conv}$ during fitting, where $r_\text{conv} = \delta \epsilon$, and $\epsilon$ is the gravitational softening of Equation \eqref{eq:NBody:softening}.}
    \label{fig:rconv4}
\end{figure}

\begin{figure}
    \centering
    \includegraphics[width=\linewidth]{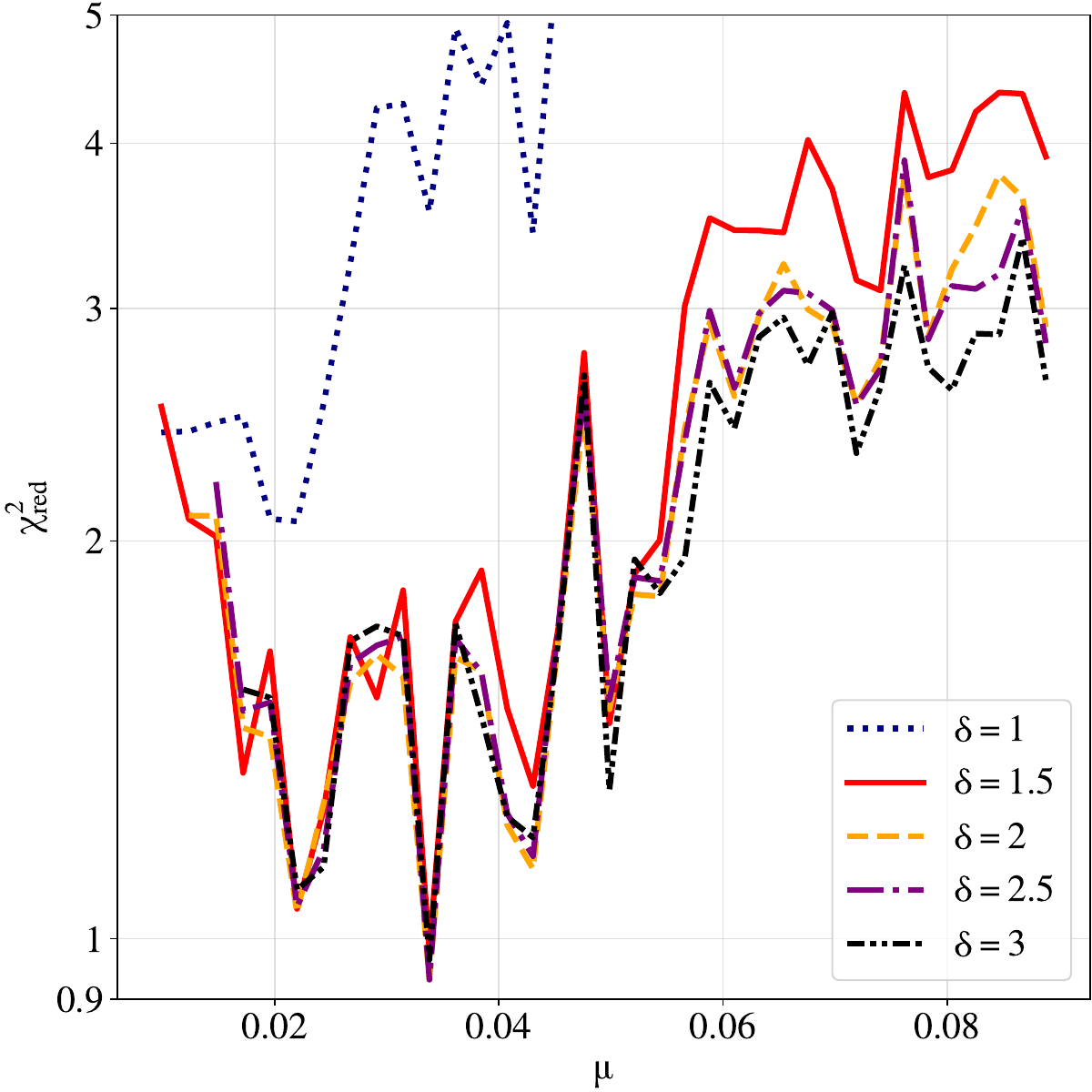}
    \caption{Values of $\chi^2_\text{red}$ for a $10^4 \text{M}_\odot$ halo and a $10^3 \text{M}_\odot$ BH grown over 4 Gyr. Every line represents a different value of $r_\text{conv}$ during fitting, where $r_\text{conv} = \delta \epsilon$, and $\epsilon$ is the gravitational softening of Equation \eqref{eq:NBody:softening}.}
    \label{fig:chiredplot}
\end{figure}

Determining the convergence radius $r_\text{conv}$ is delicate due to the nature of our simulations. Our results are of an exponential shape and if $r_\text{conv}$ is chosen too large, no significant result is fitted at all. In order to determine $r_\text{conv}$, we fit data for $r_\text{conv} = \delta \epsilon$ where $1 \leq \delta \leq 3$, in steps of 0.5, and determine at which values of $\delta$ the results remain consistent. A simulation of a $10^4 \text{M}_\odot$ halo with a $10^3 \text{M}_\odot$ BH grown over 4 Gyr has been processed for these values of $\delta$, and all found values for the proposed profile of Equation \eqref{eq:spikeProfile} are compared, together with the fits' values of $\chi^2_{red}$. See Figures \ref{fig:rconv1}, \ref{fig:rconv4}, \ref{fig:chiredplot}. The results for $\beta$ are excluded, as this parameter remains consistent over all values of $\delta$. We note that for some values, a spike was only found at later values of $\mu$. The upper boundary of the fitting was set the same for every run, and masked away the initial numerical shock wave (see Section \ref{sec:numericalshockwaves}).

We find that the results for $\delta = 2\ \&\ 2.5$ are consistent with each other, with best found values of $r_\text{sp}$ and $\gamma_\text{sp}$ falling comfortably between each others $2\sigma$-error margins. As we note the similar values of $\chi^2_\text{red}$ across these runs, we deem this to be a significant result. We have chosen to set $\delta = 2.5$ for the remainder of this article.


\section{Validation of results}
\label{sec:appendix:validation}

\subsection{Dynamical time-scales}
\label{sec:dynamicaladiabatic}

\begin{figure}
    \centering
    \includegraphics[width=\linewidth]{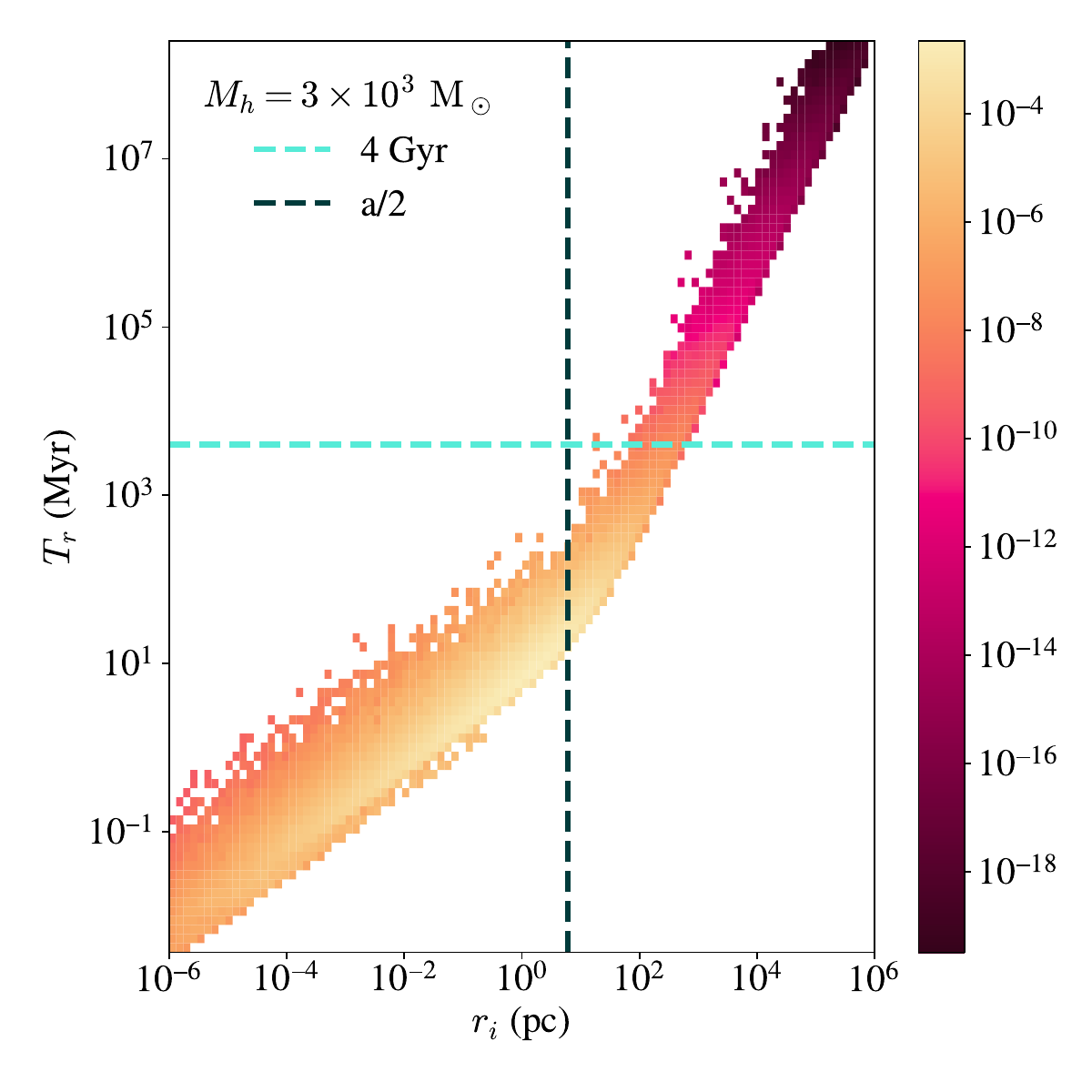}
    \caption{\label{fig:orbital_times}The distribution of radial orbital times $T_r$ as a function of the radius at which the particle was initialised $r_i$. The vertical line shows the value of $a/2$, within which all $r_\text{sp}$ lie (see Fig. \ref{fig:fitrsp}), and the horizontal line shows 4 Gyr. The colour indicates the fractional density of said orbits.}
\end{figure}

The adiabatic assumption is easily checked by comparing the orbital times of the particles in the halo with the total growth time of the central BH. By initialising a halo in a similar way to \citet{RenskPaper}, the orbital times of a Hernquist profile has been computed as a function of initial radius $r_i$, see also Figure \ref{fig:orbital_times}. This clearly shows that all particles with $r_i < a/2$ have orbital times much smaller than 4 Gyr, indicating adiabatic growth for the initial Hernquist. As the spike grows and central densities increase, these orbital times are expected to only decrease, and adiabatic growth thus holds for the entire simulation within the region of spike growth.

\subsection{Influence of resolution on fitted values}
\label{sec:appendix:influenceresolution}
In order to gauge the influence of the resolution of our simulations on the fitting of the parameters, 2 artificial systems were generated using similar binning size and Gaussian noise compared to the simulated data.\footnote{A logarithmic binning size of 0.5, and a Gaussian noise of 1\% were used.} The two artificial systems are chosen such that they represent a system with both a low $(\mu \approx 0.02)$ and high ($\mu \approx 0.1$) mass central BH. These systems were fitted twice. One using a large range of $10^{-9} \leq r \leq 10^3$, as often used in analytical studies (hereafter referred to as the full fit), and second on a small range of $10^{-3} \leq r \leq 10^0$ (hereafter referred to as the zoomed fit), comparable to our fitting range. Unique gaussian noise of 1\% is applied and fits are performed a thousand times. The average fitted values are given in Tables \ref{tab:artificialrunsLowMassMeanValues} and \ref{tab:artificialrunsHighMassMeanValues}, where we can see that there is no significant deviation between the zoomed and the full fits. 

The average standard deviations are given in Tables \ref{tab:artificialrunsLowMass} and \ref{tab:artificialrunsHighMass}. The size of the standard deviations is smaller than that of the N-body data, as the artificial data follows the fitted profile perfectly except for noise. Comparing the different results for the low mass BH in Table \ref{tab:artificialrunsLowMass}, we can see that the uncertainty on $\gamma_\text{sp}$ improves by two orders of magnitude when going from the zoomed to the full fit. The improvements for $r_\text{sp}$ and $\beta$ are more modest, indicating that these two parameters are already quite accurately fitted with a more limited range. Similar observations can be made for the high mass BH in Table \ref{tab:artificialrunsHighMass}, where the uncertainties on $r_\text{sp}$ and $\beta$ are similar for the zoomed and full fits. The uncertainty on $\gamma_\text{sp}$ improves by one order of magnitude, indicating that the zoomed fit has a less dramatic impact on the fitting uncertainty than for low-mass systems.

Concluding, we expect the fits of our systems to be very accurate for the $r_\text{sp}$ and $\beta$ parameters, with the fitting uncertainty being smaller as the mass ratio $\mu$ of the system increases. However, the resolution of these simulations is too low to determine $\gamma_\text{sp}$ with the same level of accuracy, and the effects of the fitting should be taken into account when interpreting the results.

One interesting quantity would be the necessary improvements to our N-body simulations to improve the determination of $\gamma_\text{sp}$ for the low $\mu$ system. If said system is fitted from $10^{-4.5} \leq r \leq 10^0$, an improvement factor of 11 is reached for $\gamma_\text{sp}$. Using Equation \eqref{eq:NBody:softening} to calculate the particles necessary for such a convergence radius in a $M_\text{h} = 10^{4} \text{M}_\odot$ system, we find approximately $5\times10^{10}$ particles are needed. Increasing the number of particles also decreases the error on the data points and might increase prefactor $\alpha$ of Equation \eqref{eq:NBody:softening}, and this should thus only be seen as a ways of decreasing the fitting uncertainties. Nevertheless, these are 25,000 times more particles than currently present in the simulations of said systems, and since the calculations performed by \textsc{swift} scale as $\mathcal{O}(N\log N)$, increasing the number of particles by such an amount is fully unfeasible without significant improvements to the efficiency of the code.

\begin{table}
    \centering
    \begin{tabular}{l|l|l}
    \hline
        & Zoomed-in fit & Full fit \\
        \hline
        $\langle r_\text{sp}\rangle$ & $0.050 \pm 0.001$ & $0.050 \pm 0.000$ \\
        $\langle \beta \rangle$ & $0.975 \pm 0.003$ & $0.975 \pm 0.002$ \\
        $\langle \gamma_\text{sp} \rangle$ & $-2.334 \pm 0.046$ & $-2.333 \pm 0.000$ \\
        \hline
    \end{tabular}
    \caption{The mean values of the fitted parameters of 1000 artificial systems for a low mass system where $r_\text{sp} = 0.05$, $\beta = 0.975$, $\gamma_\text{sp} = 7/3$, and ${M_\text{h} = 10^4 \text{M}_\odot}$. The error of the mean is shown in 2$\sigma$.}
    \label{tab:artificialrunsLowMassMeanValues}
\end{table}

\begin{table}
    \centering
    \begin{tabular}{l|l|l}
    \hline
        & Zoomed-in fit & Full fit \\
        \hline
        $\langle r_\text{sp}\rangle$ & $0.250 \pm 0.002$ & $0.250 \pm 0.001$ \\
        $\langle \beta \rangle$ & $0.850 \pm 0.003$ & $0.850 \pm 0.002$ \\
        $\langle \gamma_\text{sp} \rangle$ & $-2.333 \pm 0.011$ & $-2.333 \pm 0.000$ \\
        \hline
    \end{tabular}
    \caption{The mean values of the fitted parameters of 1000 artificial systems for a high mass system where $r_\text{sp} = 0.25$, $\beta = 0.85$, $\gamma_\text{sp} = 7/3$, and ${M_\text{h} = 10^4 \text{M}_\odot}$. The error of the mean is shown in 2$\sigma$.}
    \label{tab:artificialrunsHighMassMeanValues}
\end{table}

\begin{table}
    \centering
    \begin{tabular}{l|l|l|l}
    \hline
        & Zoomed-in fit & Full fit & Improvement factor \\
        \hline
        2$\sigma$ $r_\text{sp}$ & $1.03 \times 10^{-3}$ & $1.51\times10^{-4}$ & 6.82 \\
        2$\sigma$ $\beta$ & $3.41 \times 10^{-3}$ & $1.84\times10^{-3}$ & 1.85 \\
        2$\sigma$ $\gamma_\text{sp}$ & $4.57\times10^{-2}$ & $4.90\times10^{-4}$ & 93.3 \\
        \hline
    \end{tabular}
    \caption{The mean 2$\sigma$ error of the fitted parameters of 1000 artificial systems for a low mass system where $r_\text{sp} = 0.05$, $\beta = 0.975$, $\gamma_\text{sp} = 7/3$, and $M_\text{h} = 10^4 \text{M}_\odot$.}
    \label{tab:artificialrunsLowMass}
\end{table}

\begin{table}
    \centering
    \begin{tabular}{l|l|l|l}
    \hline
        & Zoomed-in fit & Full fit & Improvement factor \\
        \hline
        2$\sigma$ $r_\text{sp}$ & $2.06 \times 10^{-3}$ & $1.39\times10^{-3}$ & 1.48 \\
        2$\sigma$ $\beta$ & $3.27 \times 10^{-3}$ & $3.42\times10^{-3}$ & 0.96 \\
        2$\sigma$ $\gamma_\text{sp}$ & $1.08\times10^{-2}$ & $8.09\times10^{-4}$ & 13.3 \\
        \hline
    \end{tabular}
    \caption{The mean 2$\sigma$ error of the fitted parameters of 1000 artificial systems for a high mass system where $r_\text{sp} = 0.25$, $\beta = 0.85$, $\gamma_\text{sp} = 7/3$ and $M_\text{h} = 10^4 \text{M}_\odot$.}
    \label{tab:artificialrunsHighMass}
\end{table}

\subsection{Numerical shock waves}
\label{sec:numericalshockwaves}
An initial shock wave of particles is launched radially outward in all simulations. It is created immediately at the start of the simulation. As shown in Appendix \ref{sec:appendix:eddingtonsolutions}, this initial shock wave is present even if no BH is present or if the unmodified \textsc{swift} is used. This effect is also tested to be independent of the rate-of-growth of the central BH, the timestepping and the choice of gravitational softening. The velocity distributions of the particles are checked and behave as analytically expected. We therefore deem this effect to be due to slight imperfection in the initial conditions. We do not deem it likely this effect is due to the central BHs close range influence, as it is independent of gravitational softening. While the now missing particles will slightly change the potential of the halo, their absence is quickly overshadowed by the increasingly massive BH. Furthermore, after this shockwave has passed, the particle distribution remains unchanged, as demonstrated by the 4 Gyr old isolated haloes fully agreeing with the initial density distribution in figure \ref{fig:isolatedHalo}. This shock wave is quickly traveling, therefore easily masked away during fitting, and only disruptive for the determination of the depletion parameter $\beta$ (see Section \ref{sec:res:spikeprofile}) for the first few timesteps. We thus deem this numerical effect to not be of significant negative influence to our results.


\bsp	
\label{lastpage}
\end{document}